\documentclass[aps,physrev,reprint,superscriptaddress,longbibliography,floatfix]{revtex4-2}

\usepackage{amsmath}
\usepackage{amssymb}
\usepackage{xcolor}
\usepackage{CJK}
\usepackage[english]{babel}
\usepackage{graphicx} % Include graphics package
\begin{document}
\begin{CJK*}{UTF8}{gbsn}
%\preprint{}
\title{Modulating Low-Power Threshold Optical Bistability by Electrically Reconfigurable Free-Electron Kerr Nonlinearity}
% The Impact of Surface Charge Accumulation and Depletion on Optical Bistability


%\author{Huatian Hu (胡华天)}

\author{Huatian Hu \CJKfamily{gbsn}(胡华天)}
\email{huatian.hu@iit.it}
\affiliation{Istituto Italiano di Tecnologia, Center for Biomolecular Nanotechnologies, Via Barsanti 14, 73010 Arnesano, Italy}
\author{Gonzalo \'Alvarez-P\'erez}
\affiliation{Istituto Italiano di Tecnologia, Center for Biomolecular Nanotechnologies, Via Barsanti 14, 73010 Arnesano, Italy}
\author{Antonio Valletta}
\affiliation{Istituto per la Microelettronica e Microsistemi, Consiglio Nazionale delle Ricerche, Via del Fosso del Cavaliere 100, Rome, 00133, Italy}
%\author{...}
\author{Marialilia Pea}
\affiliation{Istituto di Fotonica e Nanotecnologie, Consiglio Nazionale delle Ricerche, Via del Fosso del Cavaliere 100, Rome, 00133, Italy}
\author{Michele Ortolani}
\affiliation{Istituto di Fotonica e Nanotecnologie, Consiglio Nazionale delle Ricerche, Via del Fosso del Cavaliere 100, Rome, 00133, Italy}
\affiliation{Department of Physics, Sapienza University of Rome, Piazzale Aldo Moro 5, 00185 Rome, Italy}
\affiliation{Istituto Italiano di Tecnologia, Center for Life Nano- and Neuro-Science, Viale Regina Elena 291, Rome, 00161, Italy}
\author{Cristian Cirac\`i}
\email{cristian.ciraci@iit.it}
\affiliation{Istituto Italiano di Tecnologia, Center for Biomolecular Nanotechnologies, Via Barsanti 14, 73010 Arnesano, Italy}
%\thanks{}


\date{\today}

\begin{abstract}
Efficient active control of optical nonlinearity has long been a critical goal for advancing nonlinear optoelectronic devices. However, this remains a significant challenge, as nonlinear susceptibilities are inherently dictated by the microscopic properties of materials.
In this work, 
we propose a microscopic mechanism to electrically reconfigure the Kerr nonlinearity by modulating the concentration of free electrons in heavily doped semiconductors under a static bias.
Our theory incorporates electrostatic and hydrodynamic frameworks to describe the electronic dynamics, demonstrating electrically tunable linear and nonlinear modulations. The power threshold of achieving optical bistability shows unprecedented tunability over two orders of magnitude, reaching values as low as 10 $\mu$W through surface charge control.
These findings offer new insights into understanding and actively controlling Kerr nonlinearities, paving the way for efficient refractive index engineering as well as the development of advanced linear and nonlinear electro-optical modulators

\end{abstract}

%\keywords{}

\maketitle
\end{CJK*}
\section{Introduction}
Refractive index engineering, which governs phase accumulation and light propagation, is central to a broad range of applications in both traditional optics and meta-optics \cite{smith2004metamaterials,galiffi2022photonics}.
Recent advances in artificial intelligence and time-varying metamaterials have highlighted the importance of active and effective refractive index modulation with low-power consumption and rapid response \cite{galiffi2022photonics}.
In this context, the optical Kerr effect \cite{boyd2008nonlinear}---a third-order nonlinear phenomenon through which light modulates a material's refractive index based on the applied intensity---has been a promising candidate. 
However, conventional Kerr nonlinearities, which arise from either intrinsic bulk nonlinear susceptibilities or thermal effects, are typically limited, respectively, in strength or speed (e.g., kHz to MHz \cite{almeida2004optical}), and thus generally unable to support high-contrast, ultrafast modulation.
%Thermal nonlinearities with strong modulation, for instance, often operate at frequencies from the kHz to MHz range. 

An effective way to overcome these challenges is by leveraging light confinement in nanoplasmonic structures.  %Specifically, plasmons, quasiparticles resulting from 
Resulting from collective free-electron (FE) oscillations, plasmons can significantly amplify the local fields and hence enhance nonlinearities in extremely confined volumes. 
Beyond the conventional nonlinear responses arising from the anharmonicity of bound electrons, the dynamics of FEs
%in the conduction band 
can generate strong and ultrafast optical nonlinearities, known as FE nonlinearities \cite{scalora2010second,ciraci2012second,scalora_electrodynamics_2020}.
Heavily doped semiconductors are particularly compelling in this context due to their ability to support exceptionally large and highly tunable optical nonlinearities \cite{lee2014giant,yu2022electrically}. 
Compared to other degenerate electron systems, such as gold \cite{scalora2010second,ciraci2012second} and indium tin oxide \cite{scalora_electrodynamics_2020,un2023electronic}, heavily doped semiconductors typically feature a lower equilibrium electron density $n_0$ ($10^{18}$ to $10^{19}$ cm$^{-3}$), {comparable to that of doped graphene \cite{cox2014electrically,rasmussen2023nonlocal},} thus typically working in the mid-infrared to THz \cite{taliercio2019semiconductor}. 
Owing to the $n_0^{-2}$ dependence of the nonlinear FE polarizability  \cite{de2021free,hu2024low}, their lower $n_0$ allows significant FE nonlinearities that can readily exceed the dielectric response from third-order nonlinear susceptibility $\chi^{(3)}$ \cite{rossetti2024origin}.
Furthermore, since FE nonlinearities act as boundary effects \cite{scalora2010second,ciraci2012second}, actively modulating the surface carrier density via field effects enables reconfigurable control \cite{de2022impact}.
Finally, unlike thermal effects that are inherently slow \cite{almeida2004optical}, FE nonlinearities from the electric polarizability of FE can be ultrafast. 
%These properties, together with the fact that semiconductors offer a well-established platform for integrated electrical tunability, as they are widely used in traditional metal-oxide-semiconductor transistors, make FE nonlinearities in heavily doped semiconductors a promising platform to actively reconfigure optical nonlinearities. 

From the theoretical point of view, FE nonlinearities can be effectively described using a microscopic, electronic-level hydrodynamic theory (HT) \cite{ciraci2012second,scalora2010second,rossetti2024origin,scalora_electrodynamics_2020,hu2024low}.
The HT is a semiclassical first-principle-level framework that can be derived from the single-particle Schr\"odinger equation in the limit of many electrons, by properly choosing the energy functionals as quantum corrections \cite{ciraci2017current,toscano2015resonance}. 
%Therefore, it can accurately describe optical nonlinearities by successfully capturing the quantum nature of FEs. 
Leveraging the HT, in a preliminary work \cite{hu2024low}, we discovered a strong FE Kerr-type nonlinearity that enables low-threshold bistability which allows two stable outputs at the same input intensity of  1 mW. 
%Despite the exceptional strength of FE Kerr nonlinearity, its real-time tunability has never been demonstrated, largely due to a lack of theoretical models to describe such an electrically reconfigurable Kerr effect. 
Despite its exceptional strength, the real-time tunability of FE Kerr nonlinearity remains unproven, mainly due to the absence of theoretical models for describing electrically reconfigurable effects.

In this work, we present a microscopic picture of the electrically tunable FE Kerr effect by implementing an electrostatics-HT coupled formalism that captures the nonlinear electron dynamics under an optical drive and field-effect gate. As a demonstration, we harness longitudinal bulk plasmons (LBPs) \cite{ruppin2001extinction} in a heavily n-doped InGaAs slab to support FE Kerr nonlinearity \cite{hu2024low,alvarez-perez_ultrahigh_2025}. 
As a nonlocal resonance above the plasma frequency, the LBP in heavily doped semiconductors enables operation from the telecom band to the mid-infrared.
In the system we propose here, LBPs are coupled to gold nanopatch antennas that function both as plasmonic cavities and electrodes for applying bias.
Notably, the LBP resonance exhibits a high sensitivity of tunability on bias, effectively functioning as a linear modulator. From a nonlinear perspective, we prove that second-order FE nonlinearities contribute to the third-order Kerr effect via a \textit{cascaded} mechanism. 
This conventionally negligible cascaded effect might be even stronger than the direct third-order process.
By incorporating the electrostatic equation, we introduce an additional degree of freedom---the static bias voltage---to manipulate FE Kerr nonlinearities. Moreover, by electrically modulating the equilibrium density $n_0$ and its gradient $\nabla n_0$, we demonstrate the ability to tune the power threshold across a 2-order-of-magnitude range, from the level of several milliwatts down to 10 $\mu$W, considering a focused beam to the diffraction limit. It can modulate both the contrast between on- and off-states and the width (read margin) of the hysteresis. 
These findings offer a microscopic understanding of Kerr nonlinearity at the nanoscale, and lay a groundwork for efficient, actively tunable linear and nonlinear devices with operating bands extending from the mid-infrared to the 1550 nm telecom band.
\section{Hydrodynamic approach and field-controlled charge modulation}
Understanding the motion of FEs is the key to decoding the optical responses of heavily doped semiconductors.
Following the HT as in Refs. \cite{de2021free,de2022impact,hu2024low}, we can model microscopic FEs driven by electric $\mathbf{E}$ and magnetic $\mathbf{H}$ fields as a fluid.  The equation of motion of these FEs can be described with two macroscopic quantities-electron density $n(\mathbf{r},t)$ and velocity $\mathbf{v}(\mathbf{r},t)$:
\begin{equation}\label{eq:HT-EqMotion}
\begin{aligned}
m_e\left( {\frac{\partial }{{\partial t}} + \mathbf{v} \cdot \nabla  + \gamma } \right)\mathbf{v} = -e({\bf{E}} + \mathbf{v} \times {\mu_0 }{\bf{H}}) - \nabla \frac{{\delta {G}[n]}}{{\delta n}}.
\end{aligned}
\end{equation}
$m_e$, $e$, $\mu_0$ are the effective mass, electron charge, and vacuum permeability respectively. The equation contains convection ($\mathbf{v}\cdot\nabla\mathbf{v}$), dissipation (phenomenological damping rate $\gamma$), Coulomb and Lorentz forces, and a quantum pressure term where an appropriate internal energy functional $G[n]$ can be selected to resolve the complex interaction in the electronic ensemble. 
The total electron density $n=n_0({\bf r})+\sum_{j}n_j(t,{\bf r})$ is comprised of the equilibrium electron density \(n_0(\mathbf{r})\) and perturbed densities (e.g., $n_j(j\omega,\mathbf{r}),j\in \mathbb{Z}$ is the order of the harmonic). Here we consider the energy functional within the Thomas-Fermi (TF) approximation, i.e., $G[n]\simeq T_{\mathrm{TF}}[n]$.
This term is essential for accurately describing nonlocal effects in plasmonics and electron dynamics \cite{ruppin2001extinction}. Additionally, more complex treatments can account for the electron spill-out effects \cite{toscano2015resonance,ciraci2016quantum}, giving rise to a slight shift of the plasmon resonance and a spatially variant $n_0$. 
For simplicity, we neglect the spill-out that allows the wavefunction of electrons to spread beyond the material's physical boundary; however, the gradient of $n_0$ can be captured as follows. 

In metal-oxide-semiconductor (MOS) structures, the electron distribution is modulated by the work function difference between the metal gate and the semiconductor, and the bias voltage $u$ on the gate. 
% When semiconductors come into contact with materials like gate oxides or metallic electrodes, $n_0(\mathbf{r})$ undergoes redistribution based on the difference in work functions between the materials.
De Luca et al. demonstrated that such a distribution can indeed have a significant impact on harmonic generation \cite{de2022impact}.  
In this regard, we consider the electrostatic equation (Eq. \ref{eq:poisson}) that is suitable for computing the semiconductors' spatially redistributed electron \(n_0({\bf r})\) and hole \(p_0({\bf r})\) densities near the interface in the presence of a band bending $\varphi$:
% In this regard, we implemented an electrostatic equation (Eq. \ref{eq:poisson}) that is suitable for describing semiconductors' spatially redistributed \(n_0({\bf r})\) near the junctions.  Therefore, a Poisson equation must be solved considering a static potential $u$, 
\begin{equation}\label{eq:poisson}
\begin{aligned}
\nabla^2\varphi=-e(p_0(\mathbf{r})-n_0(\mathbf{r})+N_D^+-N_A^-)/\varepsilon_0\varepsilon_{\rm r}.
\end{aligned}
\end{equation}
On the left-hand side of Eq. \eqref{eq:poisson} there is the Laplacian of the band bending $\varphi$, whereas the right-hand side accounts for the total charge density, including hole $p_0(\mathbf{r})$ and electron $n_0(\mathbf{r})$ densities, and ionized donor $N_D^+$ and acceptor $N_A^+$ concentrations. $\varepsilon_0\varepsilon_{\rm r}$ denotes the permittivity of the semiconductor. To solve this, we use COMSOL Multiphysics as a finite element solver for its convenience in multi-physical coupling and weak forms customization (see Supplementary Material \cite{supp} for further details). 

\begin{figure}[htbp]
\includegraphics[width=0.5\textwidth]{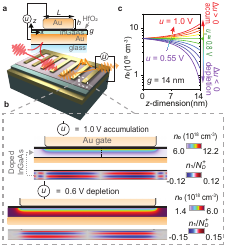}% Here is how to import EPS art
\caption{(a) The schematic of the nanopatch-InGaAs MOS configuration with external bias modulation. The height $h$ and thickness $g$ of the antenna is 50 nm and 14 nm, respectively. The length $L$ and periodicity $P_x$ can vary. (b) {The accumulation and depletion of the} equilibrium FE $n_0$ tuned by the gate bias voltage $u$, {with their normalized first-order perturbed charge density $n_1/N_{D}^+$ shown below. $n_1$ field presents the hybrid third-order LBPs in the InGaAs. The input power for calculating $n_1$ is considered as 1 mW/$\mu$m$^2$, i.e., the characteristic power level considered in the following figures. Nanopatch's $L=170$ and 210 nm for $u=1.0$ and 0.6 V, respectively, to achieve the same LBP-LSP detunings in both cases.} (c) Bias-dependent spatial distribution of $n_0$ along the dashed line in (b). }
\label{fig:1}
\end{figure}

To achieve a combination of high nonlinearity and tunability, we modified the nanopatch hybrid structure \cite{hu2024low} to align with the well-established MOS configuration \cite{jeannin2023low}. As shown in Fig. \ref{fig:1}, periodic gold stripes are designed to sit on a 2 nm HfO$_2$ \cite{zhang_vertically_2024} layer, functioning both as nanocavities \cite{baumberg2019extreme} and as gates to apply a bias for the active control. 
We choose a 2 nm HfO$_2$ for achieving an overall smaller gap thickness for a higher plasmonic field enhancement in this nanogap configuration. Larger HfO$_2$ thickness can give rise to a blueshift of the gap surface plasmon, which asks for a larger patch to compensate \cite{lassiter2013plasmonic}.
Beneath the oxide lies a thin layer of heavily doped InGaAs (14 nm thick, $N_D^+=6\times10^{18}$ cm$^{-3}$, $N_A^-=0$,  $\gamma=8.9$ ps$^{-1}$, $m_e=0.041m_0$, $\varepsilon_{\rm inf}=12$) \cite{rossetti2024origin}, which supports LBPs \cite{hu2024low}. The choice of the doping level is realistic \cite{rossetti2024origin} and its contribution to the hydrodynamic nonlinearities has been discussed in previous references \cite{rossetti2024origin,de2021free}.
Applying an external static field will tailor the FE $n_0(\mathbf{r})$  near the surface, as demonstrated by solving Eq. \eqref{eq:poisson} under different gate bias voltages $u$ (see Methods in {the Supplemental Material~\cite{supp}, including reference \cite{olmon_optical_2012}}). %We assume Ohmic contact at the heavily doped InGaAs back interface and a ``thin insulator gate'' boundary condition from COMSOL to account for the electrostatics in the insulator. %See parameters and detailed methods in Supplementary Material \cite{supp}. 

{Figure} \ref{fig:1}b shows the spatial distribution (i.e., depletion and accumulation) of $n_0$ with two different biases, 0.6 V and 1.0 V, respectively. As $u\simeq0.8$ V (Fig. \ref{fig:1}c), $n_0$ almost remains constant in space because the applied voltage balances the difference in work functions and aligns the Fermi levels (flat band condition). Therefore, we set 0.8 V as a reference gate bias $u_0$. 
In turn,  in Fig. \ref{fig:1}b, when \( u = 1 \) V (the relative bias $\Delta u = u-u_0=0.2$ V), $n_0$ near the gate increases significantly to $ 12.2 \times 10^{18} \, \mathrm{cm}^{-3}$ (accumulation condition), which is twice the value found away from the gate. In contrast, when \( u = 0.6 \) V ($\Delta u = -0.2$ V), a  depletion condition is observed, with a value of $n_0$ on the surface that is 4.3 times smaller than in the bulk. 
The spatial $n_0$ under different bias in Fig. \ref{fig:1}c proves that the charge modulation is typically a surface effect with an effective depth of the order of 10 nm. 
However, since the semiconductor thickness (14 nm) here is comparable with this depth, the surface charge modulation can have an exceptional influence on the electrical and optical performances. Such small semiconductor thicknesses combined with the small effective mass of FE typically lead to 2D confinement of free electrons and to discrete electron energy states in a quantum well
 \cite{jeannin2023low,qian2016giant}. However, it has been shown \cite{delteil2012charge} that for the very high electron density levels considered in this work, which are not typically used in quantum wells, the electromagnetic excitation spectrum is defined by the FE density parameter alone, and the thin semiconductor layer becomes a 3D plasmonic conductor with a continuum of energy levels, and not by the energy distance between discrete states as in a quantum well.

{To optimize the nonlinearity}, we focus on the LBP because it is a charge density wave in the bulk ({see first-order perturbed charge $n_1$} fields shown as Fig. \ref{fig:1}b) that overlaps with the physical dimension of InGaAs which guarantees enough active interaction volume and strong nonlinearity \cite{hu2024low}. 
As shown in Fig. \ref{fig:1}b, the induced charge $n_1$ of LBP can well overlap with the depletion layer configured by the bias, {which guarantees effective reconfiguration}. {The depletion and accumulation under biases will effectively tune the frequency of the LBP, as will be discussed in the next section. Therefore, for a fair comparison of $n_1$ between two different biases in Fig.~\ref{fig:1}b, we slightly modify the widths of the nanopatches to match the shift of LBPs ($L=170$ and 210 nm for $u=1.0$ and 0.6 V, respectively; see Supplementary Material Table II \cite{supp}). The change on the lateral dimension ($L$) has a negligible effect on the gradient of equilibrium $n_0$ on the vertical z-dimension under the patch (Figs. \ref{fig:1}b, c).}
The optical response can be described by Eq. \eqref{eq:constitutive}, the constitutive relation at the $j$-harmonic \cite{de2022impact} by adjusting the Eq. \eqref{eq:HT-EqMotion} with the relation  $\dot{\mathbf{P}}=\mathbf{J}=-ne\mathbf{v}$:
\begin{equation}\label{eq:constitutive}
 \begin{aligned}
\frac{{{\partial ^2}{{\bf{P}}_{j}}}}{{\partial {{t}^2}}} + \gamma \frac{{\partial {{\bf{P}}_{j}}}}{{\partial t}} = \frac{{{n_0(\mathbf{r})}{e^2}}}{m_e}{{\bf{E}}_{j}} +  \beta ^2\nabla (\nabla  \cdot {{\bf{P}}_{j}}) \\
-\frac{1}{3} \frac{\beta ^2 }{n_0}(\nabla  \cdot {{\bf{P}}_{j}})\nabla n_0 + {{\bf{S}}^{\rm{NL}}_{\omega_j}},
 \end{aligned}
\end{equation}
where the first-order TF-quantum pressure term has been written as $\frac{en_0}{m_e}\nabla(\frac{\delta {T}_\mathrm{TF}}{\delta n})_1 = \beta ^2\nabla (\nabla  \cdot {{\bf{P}}_{j}})-\frac{1}{3} \frac{\beta ^2 }{n_0}(\nabla  \cdot {{\bf{P}}_{j}})\nabla n_0 $. The factor $\beta(\mathbf{r})$ is related to the sound speed in the Fermi-degenerate plasma $v_{\rm F}$ \cite{grosso2013solid},  $\beta(\mathbf{r})^2=\frac{3}{5}v_F^2=2\frac{c_{\rm{TF}}}{m_e}n_0(\mathbf{r})^{2/3}$, $c_{\mathrm{TF}}=\frac{\hbar}{m_e^2}\frac{3}{10}(3\pi^2)^{2/3}$. This equation only explicitly contains the linear terms, while the nonlinear contributions %from Coulomb, Lorentz, convective forces, and quantum pressures 
are encapsulated in $\mathbf{S}^{\rm NL}_{\omega_j}$ and will be discussed below. 
%It should be noted that, due to the spatially variant $n_0(\mathbf{r})$, the factor $\beta(\mathbf{r})$ 
%and the plasma wavelength of the InGaAs, $\lambda_{\rm{p}}(\mathbf{r})= 2\pi c \sqrt{m_e\varepsilon_0\varepsilon_{\mathrm{inf}}/(n_0(\mathbf{r})e^2)}$, 
%becomes spatially dependent as well. 
By coupling Eqs. \eqref{eq:poisson}, \eqref{eq:constitutive} (assuming $\mathbf{S}^{\rm NL}_{\omega_j}=0$) with Maxwell's equation in the frequency domain, $\nabla \times \nabla \times \mathbf {E}_j - \varepsilon_{\mathrm{r}}\frac{\omega_j^2 }{c^2} \mathbf{E}_j - \mu_0\omega_j^2 \mathbf{P}_j=0,$
%\begin{equation}\label{eq:MaxwellFreq}
%\begin{aligned}
%\nabla \times \nabla \times \mathbf {E}_j - \varepsilon_{\mathrm{r}}\frac{\omega_j^2 }{c^2} \mathbf{E}_j - \mu_0\omega_j^2 \mathbf{P}_j=0,
%\end{aligned}
%\end{equation}
we can now analyze the electrical modulation of linear optical response ($j=1$). Nonlinear responses ($\mathbf{S}^{\rm NL}_{\omega_j}\neq0$) will be studied in the time domain separately later.

\section{Electrically reconfigurable linear spectrum modulation}
Figure \ref{fig:linearmodulation} illustrates the linear, electrical modulation of the plasmonic resonances by the bias $u$ within a system with $L = 184$ nm, $P_x = 280$ nm. We choose these parameters to realize a double-peak feature: plasmonic absorption enhancement as shown in Fig. \ref{fig:linearmodulation}a. As proven in Ref. \cite{hu2024low}, a slightly detuned plasmon-enhanced perfect absorption will have the optimized linear and nonlinear properties.
As $\Delta u=0$, the reflectance presents two distinct resonances at $\lambda=2.2$ and $2.4$~$\mu$m (Fig. \ref{fig:linearmodulation}a). The narrow resonance is a LBP resonance in the InGaAs layer, and the broader peak comes from the localized surface plasmon (LSP) of the nanopatch antenna \cite{hu2024low}. This LBP has been experimentally observed in various degenerate electron systems, e.g., alkali metals \cite{anderegg1971optically}, transparent conducting oxides \cite{de2018viscoelastic}, and heavily doped semiconductors \cite{moreau2024opticalexcitationbulkplasmons}. 
Enhanced by the nanopatch fundamental mode (the LSP $\simeq 2.4 \,\mu$m), this prominent hybrid LBP is an ideal platform for generating bistability with low power consumption. 

Applying a relative bias \(\Delta u\) on the antenna from -0.25 to 0.20 V can induce a significant blueshift of the LBP (Fig. \ref{fig:linearmodulation}b)---with a sensitivity of \(\frac{\Delta\lambda}{\Delta u} = 0.89\,\frac{\mu\mathrm{m}}{V}\) at the \(\Delta u = 0\). This tunability is due to the condition of charge accumulation and depletion.
For \(\Delta u > 0\), \(n_0\) tends to accumulate near the interface which leads to an effective blueshift in the plasma wavelength $\lambda_{\rm{p}}(\mathbf{r})= 2\pi c \sqrt{m_e\varepsilon_0\varepsilon_{\mathrm{inf}}/(n_0(\mathbf{r})e^2)}$. Conversely, for \(\Delta u < 0\), the resonance redshifts accordingly due to the depletion. 
Although the accumulation and depletion layers are thin (Fig. \ref{fig:1}c) and typically regarded as surface effects, they can still significantly influence plasmon resonance due to the limited thickness of the InGaAs. 
During the shift, the LBP maintains a strong interaction with light, resulting in an absorption of over 70\%.
These results reveal the sensitivity and effectiveness of linear electrical modulation, where the finding of these electrically tunable plasmons in semiconductors paves to road to wide applications in optical switches, modulators, and sensors in the infrared.
\section{Electrically Reconfigurable Free-electron Kerr nonlinearity}
We now investigate the nonlinear modulation of the optical bistability via the FE-driven Kerr effect. 
To do so, we have to solve the nonlinear problem self-consistently in the time domain to allow the field to be self-modulated by its own intensity. 
Here, as indicated by $\mathbf{S} ^{(3)}_{\omega _1}$ and $\mathbf{S} ^{(2)}_{\omega _1}$, we elaborate on the role of third- and second-order (i.e., via cascaded effects \cite{stegeman1996chi,rasmussen2023nonlocal,yang2023transformation}) FE nonlinearities respectively in Kerr effects. Fields and nonlinear sources are expanded by harmonics for an clearer comparison of different contributions:
\begin{widetext} % Allows equations to span both columns in APS template

\begin{subequations}
\label{eq:nonlinearsources}

\begin{equation}
\label{eq:SHG}
\begin{aligned}
\mathbf{S} ^{\rm (2)}_{\omega _2} &= \frac{e}{m_e} \mathbf{E}_1\nabla \cdot \mathbf{P}_1 - \frac{e\mu_0}{m_e} \dot{\mathbf{P}}_1 \times \mathbf{H}_1+ \frac{1}{en_0} (\dot{\bf P}_1 \nabla \cdot \dot{\bf P}_1 + \dot{\bf P}_1 \cdot \nabla \dot{\bf P}_1)  - \frac{1}{en_0^2} \dot{\bf P}_1 (\dot{\bf P}_1 \cdot \nabla n_0) \\
&+\frac{1}{9} \frac{\beta^2}{en_0^2}[3n_0\nabla(\nabla\cdot\mathbf{P}_1)^2-(\nabla\cdot\mathbf{P}_1)^2\nabla n_0],
\end{aligned}
\end{equation}
% First equation block
\begin{equation}
\label{eq:cascaded}
\begin{aligned}
\mathbf{S} ^{(2)}_{\omega _1} &= \frac{e}{m_e} (\mathbf{E}_1^*\nabla \cdot \mathbf{P}_2 + \mathbf{E}_2 \nabla \cdot \mathbf{P}_1^*) 
- \frac{e\mu_0}{m_e} (\dot{\mathbf{P}}_1^* \times \mathbf{H}_2 + \dot{\mathbf{P}}_2 \times \mathbf{H}_1^*)
+ \frac{1}{en_0} (\dot{\bf P}_1^* \nabla \cdot \dot{\bf P}_2 + \dot{\bf P}_2 \nabla \cdot \dot{\bf P}_1^* + \dot{\bf P}_1^* \cdot \nabla \dot{\bf P}_2 + \dot{\bf P}_2 \cdot \nabla \dot{\bf P}_1^*) \\
& - \frac{1}{en_0^2} \left[ \dot{\bf P}_1^* (\dot{\bf P}_2 \cdot \nabla n_0) + \dot{\bf P}_2 (\dot{\bf P}_1^* \cdot \nabla n_0) \right]
+ \frac{2}{9} \frac{\beta^2}{en_0^2} \left[ 3n_0\nabla \left[ (\nabla \cdot \mathbf{P}_2) (\nabla \cdot \mathbf{P}_1^*) \right] - (\nabla \cdot \mathbf{P}_2) (\nabla \cdot \mathbf{P}_1^*) \nabla n_0 \right],
\end{aligned}
\end{equation}

% Second equation block
\begin{equation}
\label{eq:direct}
\begin{aligned}
\mathbf{S} ^{(3)}_{\omega _1} & = -\frac{1}{e^2n_0^2} [  \nabla \cdot \mathbf{P}_1^*\dot{\bf P}_1\nabla \cdot \dot{\bf P}_1 +  \nabla \cdot \mathbf{P}_1\dot{\bf P}^*_1\nabla \cdot \dot{\bf P}_1 +  \nabla \cdot \mathbf{P}_1\dot{\bf P}_1\nabla \cdot \dot{\bf P}_1^*
+\nabla \cdot \mathbf{P}_1^*\dot{\bf P}_1\cdot \nabla \dot{\bf P}_1+\nabla \cdot \mathbf{P}_1\dot{\bf P}_1^*\cdot \nabla \dot{\bf P}_1\\
&+\nabla \cdot \mathbf{P}_1\dot{\bf P}_1\cdot \nabla \dot{\bf P}_1^*
+2\dot{\mathbf{P}}_1^*\dot{\mathbf{P}}_1\nabla \nabla \cdot\mathbf{P}_1+\dot{\mathbf{P}}_1\dot{\mathbf{P}}_1\nabla \nabla \cdot\mathbf{P}_1^*] \\
&+\frac{2}{e^2n_0^3}[(\nabla \cdot \mathbf{P}_1^* )\dot{\mathbf{P}}_1(\dot{\mathbf{P}}_1\cdot \nabla n_0)+(\nabla \cdot \mathbf{P}_1 )\dot{\mathbf{P}}_1^*(\dot{\mathbf{P}}_1\cdot \nabla n_0)  +(\nabla \cdot \mathbf{P}_1 )\dot{\mathbf{P}}_1(\dot{\mathbf{P}}_1^*\cdot \nabla n_0)] \\
&-\frac{4}{27}\frac{\beta^2}{e^2n_0^3}[\frac{3}{4}n_0\nabla[\nabla \cdot \mathbf{P}_1^*(\nabla \cdot \mathbf{P}_1)^2]-\nabla \cdot \mathbf{P}_1^*(\nabla \cdot \mathbf{P}_1)^2\nabla n_0],
\end{aligned}
\end{equation}
% Third equation block

\end{subequations}

\end{widetext}

\begin{figure}[htbp]
\includegraphics[width=0.4\textwidth]{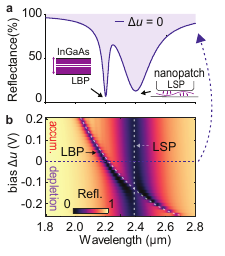}% Here is how to import EPS art
\caption{Linear spectral modulation as a function of the applied bias (b) with one specific case (a), $\Delta u=0$. The sketches in (a) represent the resonances of LBP due to the nonlocality in the InGaAs and LSP in the nanopatch cavity.} 
\label{fig:linearmodulation}
\end{figure}

\noindent where ${\bf{S}}_{\omega_2}^{\rm{NL}}={\bf{S}}_{\omega_2}^{\rm{(2)}}$ generates the second harmonic ($\omega_2=2\omega_1$) waves and the Kerr nonlinear sources can be broken down according to the orders of the processes ${\bf{S}}_{\omega_1}^{\rm{NL}}={\bf{S}}_{\omega_1}^{(2)}+{\bf{S}}^{(3)}_{\omega_1}$. Here, ${\bf{S}}_{\omega_1}^{(3)}$ denotes the free-electron Kerr nonlinear source due to the direct process through a third-order nonlinearity  (i.e., denoted by the superscript (3)). It only contains the field at $\omega_1$. ${\bf{S}}_{\omega_1}^{(2)}$ is the second-order (i.e., denoted by the superscript (2)) source that gives rise to the Kerr nonlinearity (i.e., denoted by the subscript $\omega_1$) by a cascaded effect. 
It is determined by the field at second harmonic $\omega_2$ (e.g., $\mathbf{P}_2$ and $\mathbf{E}_2$) and the conjugated fields from fundamental frequency $\omega_1$  (e.g., $\mathbf{P}_1^*$ and $\mathbf{E}_1^*$). 
{This strong second-order effect originates from the nonlocality of the field, which was also found in other FE systems such as highly-doped graphene~\cite{rasmussen2023nonlocal}, where third-harmonic fields can be generated through cascaded effects.}
The ``$\cdot$" and ``$*$" on the variables denote the time derivative and complex conjugate, respectively.

Note that Eqs. \eqref{eq:SHG}-\eqref{eq:direct} contain contributions from the gradient of the equilibrium density, i.e., $\nabla n_0$, which may have a significant influence on the nonlinear sources near the interface due to the abrupt change of $n_0$ (Fig. \ref{fig:1}c). 
Beyond this, nonlinear sources contain terms proportional to $n_0^{-1}$ and $n_0^{-2}$. It suggests an electrically reconfigurable Kerr effect controlled by bias $u$ that a depletion of $n_0$ will significantly enhance the nonlinearity, and conversely, an accumulation may reduce it. 
As a first note, here we have derived a full picture of hydrodynamic Kerr nonlinearity (Eqs. \eqref{eq:cascaded},\eqref{eq:direct}) under bias ($u$ may contribute via $n_0(\mathbf{r})$) and are ready to implement them once coupled with the time-domain wave equation for the vector potential $\mathbf{A}^{(j)}(\mathbf{r},t)$ at $j$-th harmonic:
\begin{equation}\label{eq:waveequationtd}
\nabla \times \nabla \times \mathbf{A}_{j}+  \frac{\varepsilon_{\mathrm{r}}}{c^2}\frac{\partial^2 \mathbf{A}_{j}}{\partial t^2}+\mu_0\frac{\partial }{\partial t}(\mathbf{P}_{j}+\mathbf{P}^{\mathrm{NL}}_{\mathrm{d}})=0.
\end{equation}
%It is important to note that we deliberately separate the harmonics of the fields for a direct comparison between direct $\mathbf{S} ^{(3)}_{\omega _1}$ and cascaded $\mathbf{S} ^{(2)}_{\omega _1}$ nonlinear channels. 
Here, the electromagnetic fields are connected to $\mathbf{A}_j$ by $\mathbf{E}_j= -\partial \mathbf{A}_j/\partial t$ and $\mu_0 \mathbf{H}_j=\nabla \times \mathbf{A}_j$. The Kerr nonlinearity from the dielectric can be considered with a polarization $\mathbf{P}^{\mathrm{NL}}_{\mathrm{d}}=3\varepsilon_0\chi^{(3)}|{\bf E}|^2{\bf E}$ with the susceptibility $\chi^{(3)}=1.6\times 10^{-18}$ m$^2/$V$^2$  \cite{boyd2008nonlinear}. 
%where the independent variables $(\mathbf{r},t)$ were omitted in notations for simplicity; 

\begin{figure}[htbp]
\includegraphics[width=0.35\textwidth]{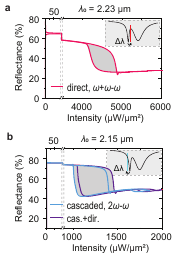}% Here is how to import EPS art
\caption{Comparison of the hysteresis resulted from different nonlinear sources: dielectric $\chi^{(3)}$ plus hydrodynamic terms (a) with the direct process only (red), (b) with the cascaded process only (cyan), and their combination (purple). Colorful lines in insets mark the laser wavelengths $\lambda_0$ which are detuned from the resonance with $\Delta\lambda$.}
\label{fig:dirvscas}
\end{figure}

After solving the system of equations given by Eqs. \eqref{eq:poisson}-\eqref{eq:waveequationtd} based on the nanopatch of Fig. \ref{fig:linearmodulation}a,  here we compare the contributions of Kerr nonlinearity from the direct and cascaded process. 
As shown in Fig. \ref{fig:dirvscas}a, the input wavelength is red-detuned compared with the LBP resonance, i.e., $\lambda_0=2.23$ $\mu$m, the LBP mode $\lambda=2.2 \,\mu$m, detuning $\Delta\lambda=\lambda_0-\lambda=30$ nm. The Kerr effect through a direct third-order process ($\mathbf{S} ^{(3)}_{\omega _1}$) will form bistability with a intensity threshold of 5 mW/$\mu \rm m^2$. 
In contrast, the cascaded process, which includes difference-frequency contributions arising from the second-order nonlinearity ($\mathbf{S} ^{(2)}_{\omega _1}$), results in a significantly lower threshold of approximately 1.5 mW/$\mu \rm m^2$ for blue-detuned excitation ($\Delta\lambda=-50$ nm), accompanied by a wider hysteresis (Fig. \ref{fig:dirvscas}b, cyan curve). 
Two hysteresis loops are observed in the reflectance (Fig. \ref{fig:dirvscas}b) due to the existence of two resonances, i.e., LBP and LSP on the lower-energy side of the input $\lambda_0$. 
Interestingly, these results suggest that the direct process is a Kerr effect with positive effective susceptibility that will redshift the resonance ($n_{2\rm Kerr}>0$, $n_{2\rm Kerr}$ is nonlinear refractive index) while the cascaded process can be a ``negative" that reshapes the resonance to the blue side ($n_{2\rm Kerr}<0$). {Such a negative $n_{2\rm Kerr}$ has also been observed in graphene FE nonlinear systems \cite{jelver2024nonlinear}.}
%Since the cascaded effect involves the field at second harmonics, e.g., $\mathbf{E}_2,\mathbf{P}_2 $, etc., the relative phase difference between the pump $\lambda_0/2$ and the higher-order plasmon resonance near $\lambda_0/2$ can determine the polar of the nonlinearity. 

Since the direct and cascaded Kerr effects exert opposite dephasing in this specific structure, combining both effects, $\mathbf{S} ^{(2)}_{\omega _1}+\mathbf{S} ^{(3)}_{\omega _1}$, for direct comparison can be useful. We observe that the hysteresis (purple curve in Fig. \ref{fig:dirvscas}b) shifts towards a higher intensity threshold, while its overall shape remains unchanged. 
These three scenarios, i.e., direct (red), cascaded (cyan), and their combination (purple), reveal for the first time that the cascaded contributions, a second-order process, can even dominate in the FE Kerr effect. %The former depends on the second-order FE nonlinearity, while the latter is a third-order effect.
{This offers a channel to engineer the FE Kerr nonlinearity, for instance, by designing multi-resonant antennas that can enhance both harmonics~\cite{noor2020mode}, thereby enhancing the cascaded nonlinear response as well for a stronger nonlinearity~\cite{rasmussen2023nonlocal}.}
%Interestingly, this observation aligns with prior studies on harmonic generation from hydrodynamic sources in heavily doped semiconductors, where cascaded FE nonlinearities can also be more efficient than direct ones \cite{de2021free}. 
In addition, the FE nonlinearity due to both cascaded and direct processes can be dominant over traditional {local} $\chi^{(3)}$, since the latter has been proven to be much weaker than the direct FE Kerr effect \cite{hu2024low}. {An interesting contrast is that the cascaded effect in highly doped graphene appears less prominent than the local third-order nonlinear response in many cases~\cite{rasmussen2023nonlocal}, whereas in heavily doped semiconductors the cascaded effect can dominate, as also discussed in Refs.~\cite{de2021free,yang2023transformation}.
}

\begin{figure*}[tbp]
\includegraphics[width=0.8\textwidth]{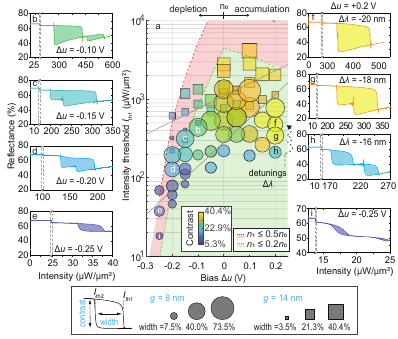}% Here is how to import EPS art
\caption{Low-power threshold optical bistabilities electrically modulated by static bias. (a) Statistics of the power threshold for achieving bistability as a function of bias voltage and InGaAs slab thickness ($g = 8, 14$ nm). See statistical data for $g = 11$ nm in \cite{supp}. The sizes (colors) of the markers represent the width (contrast) of the hysteresis. Left panels (b-e) show hysteresis loops of four specific examples with different $\Delta u$ marked in (a). The right panels (f-h) show a hysteresis-merging mechanism of the system with the same bias but different detunings. (i) shows the hysteresis with the lowest threshold collected in the statistics.}
\label{fig:statistic}
\end{figure*}

Finally, we investigate how the bias field tunes the optical bistability. 
In Fig. \ref{fig:statistic}a we present a statistics study for the optical bistabilities % generated by the nanopatch-InGaAs hybrid systems 
under different biases $\Delta u$, and detunings $\Delta\lambda$. Specific examples of $\Delta u$ (Figs. \ref{fig:statistic}b-e) and $\Delta\lambda$ (Figs. \ref{fig:statistic}f-h) dependencies are discussed. 
We investigate three InGaAs thicknesses: 8 nm, 11 nm (Fig. S3 \cite{supp}), and 14 nm. To fairly compare the strength of Kerr nonlinearity under different biases, in every scenario of Fig. \ref{fig:statistic}, we keep the detunings of the LBP and LSP the same as that shown in Fig. \ref{fig:linearmodulation}a. See details in Fig. S2 \cite{supp}. 
There are three crucial charactetristics of a bistable hysteresis: modulation depth (contrast), hysteresis width, and the power threshold. The contrast influences the fidelity, the width guarantees the writing and reading margins of optical memory, and the threshold determines the power consumption. 
As shown in the legend, we use the size of the scatters to indicate the hysteresis's width, \( w = 2\frac{I_{\rm th1}-I_{\rm th2}}{I_{\rm th1}+I_{\rm th2}} \). \( I_{\rm th1} \) and \( I_{\rm th2} \), defined at the half-maximum of the hysteresis, represent the thresholds for switching the states off and on.
Then, the contrast is defined as the difference in reflectance between the maximum and minimum of the two stable states, which is indicated by the color of the scatters. 
The green and red shades mark the regions where the perturbed electron density $n_1$ is smaller than 20\% and 50\% of the minima of $n_0(\mathbf{r})$ under different biases, respectively. 
{Note that the thresholds of 20\% and 50\% are not strict limits, they serve as indicative reference for the limit set by the perturbative approach} in the Taylor expansion ($n_1/n_0\ll 1$) which is the assumption \cite{de2021free} while deriving the FE hydrodynamic nonlinearities (Eq. \eqref{eq:cascaded}-\eqref{eq:SHG}). 
With these quantitative metrics, we can now assess our idea of tuning nonlinearity by a bias. 

In Fig. \ref{fig:statistic}a, the squares (the InGaAs layer thickness $g=14$ nm) are generally distributed above the spheres ($g=8$ nm), indicating that they require higher intensities to generate bistable states. Diamonds ($g=11$ nm) are apparently in between them, shown in Fig. S3 \cite{supp}.
While a larger thickness can diminish the field confinement in the nanopatch, it can also increase the interaction length, maintaining a decent modulation contrast (indicated by the colors of the markers).
As we reduce the InGaAs thickness from 14 nm to 8 nm, we can observe a significant drop in the threshold $I_{\rm th1}$, accompanied by an increase in the hysteresis width, as indicated by the generally larger sizes of the circles than the squares.
The contrast of $g=8$ nm systems can still be pronounced due to the compensation of the field enhancement over the loss of interaction dimension. 
More importantly, as shown by the gray guide-to-eye lines, the threshold for systems with smaller gap thickness (\(g=8\) nm) is more sensitive to the bias voltage compared to those with \(g=14\) nm. 
This high sensitivity can be attributed to the larger ratio of the charge modulation depth to the total material thickness.

Whereas $g$ is an unreconfigurable parameter, we hereby study the dependence of the power threshold and modulation contrast on electrical tuning $u$.
As illustrated by the colors and vertical-axis positions of the markers in Fig.\ref{fig:statistic}a, the charge accumulation ($\Delta u>0$) increases the contrast as well as the threshold, respectively. 
Conversely, charge depletion ($\Delta u<0$)  has the opposite effect, reducing them. 
The power threshold drops because the depletion (lower $n_0$) will strengthen the $\mathbf{S}^{\rm NL}_{\omega_1}$ due to the factors of $n_0^{-1}$, $n_0^{-2}$ and the $\nabla n_0$ near the surface. 
Meanwhile, the depletion decreases the number of electrons that may participate in the absorption, which decreases the contrast as a trade-off. Charge accumulation has the opposite effect. 

On the other hand, as shown by the red and green shades in Fig. \ref{fig:statistic}a, both charge depletion and accumulation may decrease the limit of maximum excitation intensity. 
Notably, the nonlinear process will saturate once as most of the FE are excited ($n_1\simeq n_0$). 
Thus, charge depletion reduces $n_0$, causing quicker saturation, while accumulation enhances the electric field, increasing polarizability (higher \( n_1 \) for the same input field) and facilitating easier saturation.
Based on this, for each $\Delta u$, we increase the $\Delta\lambda$ to probe the best bistable behaviors (large contrast and width) under the strongest possible excitation within the saturation restrictions: $n_1(\mathbf{r})\leq0.2$ to $0.5n_0(\mathbf{r})$. 
As indicated by three specific cases when $\Delta u = 0.2$ V in the Figs. \ref{fig:statistic}f-h, tuning $\Delta \lambda$ from 16 nm to 20 nm can gradually merge the two hysteresis loops (i.e., from the LBP and LSP, respectively) into one. 
This allows engineering the modulation contrast as well as the hysteresis width. With a larger detuning, $\Delta \lambda = 18$ and 20 nm, the modulation depth is greatly enhanced by 2 times compared with $\Delta\lambda=16$ nm case, while $\Delta \lambda = 18$ nm case may have a slightly larger width than $\Delta \lambda = 20$ nm. 
The detuning $\Delta\lambda$ has long been recognized as a tool for modulating the hysteresis width \cite{nakarmi2014analysis}. 
It can also serve as an unprecedentedly interesting method for designing bistability via merging hysteresis loops from multiple modes. 
It offers the freedom to engineer both the width and contrast of the hysteresis.

Figures \ref{fig:statistic}b-d exemplify a few hysteresis loops (whose labels are marked in Fig. \ref{fig:statistic}a),  with different charge depletion by $\Delta u$ from -0.10 to -0.25 V close to the saturation limit. 
As $n_0$ is more depleted, the intensity threshold drops from 500 to 35 $\rm \mu W/\mu m^2$ with a lower modulation contrast down to 10\%. 
The minimal power threshold achieved near the 20\% saturation limit (Fig.\ref{fig:statistic}i) is 17.5 $\rm \mu W/\mu m^2$, yet with a 5\% contrast. 
In summary, the electrical tuning of the FE Kerr nonlinearity always requires a balance between power consumption (threshold) and modulation contrast. 
To pursue an exceptionally low power consumption, one has to achieve efficient charge depletion through a bias. 
On the other hand, for a stronger modulation with high fidelity, one needs charge accumulation. Notably, even with charge accumulation, the power threshold remains as low as the milliwatt level (considering a diffraction-limit beam size) which is exceptionally efficient. 
In addition, judiciously designing the multimode systems and engineering the bistabilities by merging two or more hysteresises via tuning $\Delta\lambda$ can be an interesting tool for memory device design. 

{
With the relatively strong excitation field used in our study, heating of the free electron gas may become relevant~\cite{jelver2024nonlinear}. However, the Thomas-Fermi functional used in our work was derived with zero electron temperature \cite{grosso2013solid}, and incorporating finite-temperature corrections into our nonlinear hydrodynamic framework in a self-consistent way is currently non-trivial. On the other hand, methods based on the Boltzmann equation~\cite{un2023electronic} or quantum mechanical approaches such as the random phase approximation (RPA)~\cite{cox2014electrically,jelver2024nonlinear} could face challenges when applied to complex geometries. A more comprehensive theoretical treatment remains an open direction for future work.
That said, we can still qualitatively discuss the effects of electron heating-such as spectral broadening from electron-electron and electron-phonon collisions-by phenomenologically increasing the damping rate $\gamma$. As we detailed in our previous work (Ref. \cite{hu2024low}, SI Section 4, Fig. S3), this broadening would degrade the nonlinear performance by raising the threshold and narrowing the hysteresis width. 
Nonetheless, with a damping rate as high as $\sim3\gamma$, the bistability threshold stays around 10 mW$\cdot\mu$m$^{-2}$. Other, more sophisticated damping mechanisms-such as nonlocal or viscoelastic and nonlinear dampings-have been discussed in detail in our recent work \cite{alvarez-perez_ultrahigh_2025}, and have shown the robustness of the FE Kerr nonlinearity.}

{Finally, an experimental implementation is feasible by using InGaAs epitaxial layers co-doped in the chemical vapor deposition (CVD) chamber up to $6 \times 10^{18}$ cm$^{-3}$~\cite{rossetti2024origin}. Thicknesses of the doped InGaAs layers around $g \sim 14$ nm are routinely achieved by CVD. The most critical component is the gate oxide layer made of Hafnium oxide: it has been proven that it can hold high electric fields up to 20 MV/cm\cite{sire2007statistics}, which is the maximum value foreseen in this paper with a 2 nm thick oxide \cite{zhang_vertically_2024} and up to 4 V gate bias voltage. Milder scenarios with gate oxide thicknesses of the order of 10 nm can also be practically relevant. The gate contact can be made, for example, of a Ti/Au bilayer (5/45 nm to get a total $h = 50$ nm) and shaped as a 1-dimensional periodic grating (nanopatch) through electron-beam lithography, electron-beam deposition, and lift-off. 
Besides, the semiconductor device can naturally experience carrier depletion due to band bending even at zero bias, see low $u$ cases in Fig.~\ref{fig:1}c. Therefore, modulation effects are expected even with a slight bias variation from zero, indicating that this scheme remains effective in the low-field regime.}

\section{Conclusion}
In conclusion, we have proposed a mechanism to electrically modulate the nanoscale Kerr nonlinearity and optical bistability. 
By incorporating a microscopic HT with the carrier transport dynamics of semiconductors, we demonstrate that a static field can efficiently modulate both linear and nonlinear optical responses by spatially varying the FE equilibrium density.
%FE Kerr nonlinearities originating from second and third-order hydrodynamic nonlinear sources were compared and discussed, demonstrating that the cascaded Kerr nonlinearity from the second-order effects can surpass the direct third-order contribution.
Varying biases, a charge depletion will significantly decrease the power threshold of the bistability from thousands of down to tens of $\rm \mu W/\mu m^2$. On the contrary, charge accumulation will give rise to a wide ($> 70\%$) and high-contrast ($\sim$40\%) bistability with a still decent threshold of several $\rm mW/\mu m^2$. 
Moreover, we unveil that the often-neglected cascaded contributions in the Kerr effect can also be dominant in this system.
Our findings open a viable avenue for active ultrafast nonlinear devices with low power consumption.

\section*{ACKNOWLEDGMENTS}
We thank Raffaele Colombelli for the helpful discussions. We acknowledge funding from the European Innovation Council through its Horizon Europe program with Grant Agreement No. 101046329.
Views and opinions expressed are those of the authors only and do not necessarily reflect those of the European Union. Neither the European Union nor the granting authority can be held responsible for them.
\section*{DATA AVAILABILITY}
The data that support the findings of this article are openly available \cite{das}.
\bibliography{apssamp.bib}

\end{document}